\documentclass[superscriptaddress,twocolumn,showpacs,preprintnumbers,amsmath,amssymb,prl]{revtex4}
\usepackage{epsfig}
\newcommand{\bm}[1]{ \mbox{\boldmath $#1$}  }
\newcommand{\ud}{\mathrm{d}}
\begin{document}

\title{Stability of a BEC with Higher-order Interactions near a 
  Feshbach Resonance}

\author{N.T. Zinner}
\affiliation{ Department of Physics, Harvard University, Cambridge,
  Massachusetts 02138, USA}

\author{M. Th{\o}gersen}
\affiliation{ Department of Physics and Astronomy, University of
  Aarhus, DK-8000 Aarhus, Denmark}

\date{\today}

\hyphenation{Fesh-bach}

\begin{abstract}
  We consider the stability of an ultracold trapped Bose-Einstein
  condensate near a Feshbach resonance. Using a modified
  Gross-Pitaevskii equation that includes higher-order terms and a
  multi-channel model of Feshbach resonances, we find regions where
  stability can be enhanced or suppressed around experimentally
  measured resonances. We suggest a number of ways to probe the
  stability diagram.  Using scattering length zero-crossings huge
  deviations are founds for the critical particle number. Effects are
  enhanced for narrow resonances or tighter
  traps.  Macroscopic tunneling of the condensate is another possible
  probe for higher-order interactions, however, to see this requires
  very narrow resonances or very small particle numbers.
\end{abstract}
\maketitle

\paragraph*{Introduction}
The stability of Bose-Einstein condensates (BECs) in ultracold alkali
gases is determined by the sign of the scattering length $a$
\cite{dalfovo1999}. For $a<0$, one has effectively attractive
interactions and the condensate will collapse to a dense state when
the number of condensed particles $N$ is larger than a critical number
$N_c$ \cite{ruprecht1995,baym1996,dalfovo1999,gammal2001}. This has
been beautifully demonstrated in experiments with $^{7}$Li
\cite{bradley1997}, $^{87}$Rb \cite{roberts2001,donley2001}, and
recently with a dipolar $^{52}$Cr BEC \cite{lahaye2008}. The findings
indicate that the theory based on the surprisingly simple
Gross-Pitaevskii (GP) equation can reproduce and describe most
features of the experiments.

The GP equation includes two-body terms through a contact interaction
which is parametrized by $a$. This is equivalent to a Born
approximation but with an effective coupling that is obtained by
replacing $a_{born}$ by the physical scattering length $a$. However,
higher-order terms in the expansion of the phase shifts at low
momenta, determined by the effective range $r_e$, the shape parameter
etc., give corrections to the simple GP equation. In this paper we
explore the influence of the effective range term on the
properties of a BEC. In particular, we show that the critical number
of condensed atoms depends strongly on the higher-order scattering
term when the scattering length approaches zero (zero-crossing). We
also show how the macroscopic quantum tunneling (MQT) rate, in which
the entire BEC tunnels as a coherent entity, can be modified for small
condensate samples.

The considered effects depend on a combination of $a$ and $r_e$ which
yield different behavior for wide and narrow Feshbach resonances.
Recent measurements on $^{39}$K found many both wide and narrow
resonances which allow for tuning of $a$ over many orders of magnitude
\cite{roati2007,errico07}. We therefore consider a selected example
from $^{39}$K in order to elucidate the general behavior for realistic
experimental parameters.

The paper is organized as follows: We first introduce the modified GP
equation with effective range dependence. Using both a variational and
numerical approach we find a phase diagram describing the stability of
the BEC.  We then consider a Feshbach resonance model
including effective range variations.  The behavior of the critical
particle number near a scattering length zero-crossing is derived.  We
discuss MQT and show numerically how the rate is modified. We finally
discuss other possibilities for probing the higher-order interactions.

\paragraph*{Modified GP Equation}
We assume that the condensate can be described by the GP equation and
we focus on the $a<0$ attractively interacting case.  Since we are
interested in the ultracold regime, where the temperature is much
smaller than the critical temperature for condensation, we adopt the
$T=0$ formalism. In order to include higher-order effects in the
two-body scattering dynamics, we use the modified GP equation derived
in \cite{pet07} for which the equivalent energy functional is
\begin{align}\label{efunc}
  E(\Psi)=\int \ud{\bm r} 
  &\left[\frac{\hbar^2}{2m}|\nabla \Psi|^2+V(\bm r)
  |\Psi|^2\right.&\nonumber\\
  &\left.+\frac{U_0}{2}\left(|\Psi|^4 +g_2|\Psi|^2
      \nabla^2|\Psi|^2\right) \right],&
\end{align}
where $m$ is the atomic mass, $V$ is the external trap,
$U_0=4\pi\hbar^2 a/m$, and $g_2=a^2/3-ar_e/2$ with $a$ and $r_e$ being
the $s$-wave scattering length and effective range, respectively.

\begin{figure}[htb]
  \epsfig{file=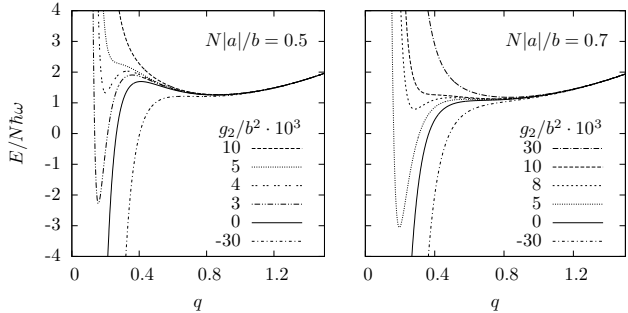,clip=true,scale=0.70}
  \caption{Energy of a BEC with fixed $N\lvert a\rvert/b$ as function
    of the variational parameter $q$, i.e. the size of the condensate.
    The higher-order interaction term $g_2$ modifies the height and
    shape of the barrier.}
  \label{fig1}
\end{figure}

We are interested in the stability properties of the ground-state and
we therefore perform a variational calculation using the mean-field
trial wavefunction
\begin{equation}\label{trial}
  \Psi(r)
  =\frac{\sqrt{N}}{\pi^{3/4} \sqrt{(qb)^3}}\exp(-\frac{r^2}{(qb)^2}),
\end{equation}
where $q$ is the dimensionless variational parameter and
$b=\sqrt{\hbar/m\omega}$ is the trap length. The normalization is
\mbox{$N=\int \ud{\bm r}|\Psi(r)|^2$}. For simplicity we only consider
isotropic traps with $V(r)=\tfrac{1}{2}m\omega^2r^2$.  However, the
effects found should hold for deformed traps as well (along the lines
of the analysis in \cite{ueda1998}).  The variational energy is
\begin{equation}\label{energy}
  \frac{E(q)}{N\hbar\omega}
  =\frac{3}{4}q^2+\frac{3}{4}\frac{1}{q^2}+\frac{1}{\sqrt{2\pi}}\frac{N|a|}{b}
\left(-\frac{1}{q^3}
  +3\frac{g_2}{b^2}\frac{1}{q^5}\right).
\end{equation}
In Fig.~\ref{fig1} we plot $E(q)$ for different parameters. As shown,
there are many possibilities for $g_2 \neq 0$, including stable,
unstable, and metastable systems. We see that the $g_2$ term modifies
the barrier for $N|a|/b=0.5$, implying that tunneling rates will be
altered.  For $N|a|/b=0.7$, the $g_2=0$ case has no barrier at all,
and here the $g_2$ term can in fact produce a small barrier on its
own. The $q^{-5}$ dependence of the term means that the effect is
small. However, the plot clearly shows that a new stability analysis
is needed.  In addition to the variational approach we have
numerically solved the full time-independent GP equation corresponding
to eq.~\eqref{efunc}.

\begin{figure}[tb]
  \epsfig{file=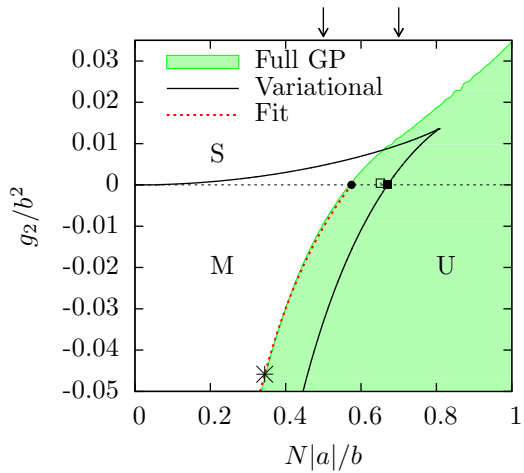,clip=true,scale=1.0}
  \caption{(color online) Phase stability diagram of a BEC with
    higher-order interactions. The solid line obtained from the
    variational ansatz, eq.~\eqref{trial}, divides the stable~(S),
    metastable~(M), and unstable~(U) regions.  The white region
    indicates where stationary solutions exists for the full GP
    equation. Filled points show the well-known $g_2=0$ results.
    Arrows correspond to the values in Fig.~\ref{fig1}.  The dashed
    line is the fit in eq.~\eqref{eq:Nc-fit}. Two specific $^{39}$K
    values are chosen for zero-crossing (cross) and MQT rates (open
    square) calculations (see text).}
  \label{fig2}
\end{figure}

\paragraph*{Phase Stability Diagram}
To determine the ground-state stability one looks for the vanishing of
the barrier towards $q=0$. For $g_2=0$ eq.~\eqref{energy} leads to
$N_c |a|/b\approx 0.671$ \cite{dalfovo1999}. The full integration of
the GP equation gives $N_c |a|/b\approx k_0$, $k_0=0.5746$
\cite{ruprecht1995,gammal2001}. These values are indicated by filled
points in Fig.~\ref{fig2}.

In the general $g_2\ne0$ case we first take the variational energy
eq.~\eqref{energy} and solve for multiple roots of $\ud E(q)/\ud q$.
The resulting ``phase diagram'' is plotted in Fig.~\ref{fig2} (solid
line). In the upper left region (S) we have complete stability of the
condensate; only one minimum exists at large $q\sim 0.8$ (see
Fig.~\ref{fig1}) and the potential goes to plus infinity at small $q$.
In the metastable region (M) a barrier and a minimum exists for large
$q$; for $g_2>0$ another minimum exists at small $q\lesssim 0.2$,
while for $g_2<0$ the potential goes to minus infinity.  In the
unstable region (U) the barrier has vanished; the potential either has
a minimum at small $q$ ($g_2>0$) or no minimum at all ($g_2<0$).  In
the variational approach the stable and unstable regions are connected
via the upper right part of Fig.~\ref{fig2}. Going from (S) to (U)
corresponds to an adiabatic change, where the macroscopic ($q\sim 1$)
minimum is transferred to a microscopic ($q\ll 1$) high-density
minimum.

Next, we numerically solve the full the GP equation. The stationary
(white) and non-stationary (shaded) regions are shown in
Fig.~\ref{fig2}. Since both stable and metastable solutions are
considered stationary, the white regions cover both (S) and (M). Both
our variational and numerical results agrees with the known $g_2=0$
results, and have similar behavior for small and negative $g_2/b^2$.

\paragraph*{Feshbach Resonance Model}
In order to predict effects of the higher-order term, we need a
realistic model for $g_2$. Since $g_2/b^2$ is the relevant parameter,
and the trap length, $b$, is usually orders of magnitude larger than
atomic scales, it is necessary to look for divergences of $g_2$. Since
$g_2$ depends on $a$ and $r_e$, a Feshbach resonance is the obvious
mean. The standard single-channel models are inadequate since only $a$
is considered.  We therefore use a multi-channel model \cite{bruun05},
which describes both $a$ and $r_e$ as a function of resonance position
$B_0$, width $\Delta B$, magnetic moment difference between the
channels $\Delta \mu$, and the background scattering length $a_{bg}$.
In \cite{bruun05}, only the regime $|B-B_0|\ll \Delta B$ was of
interest, yielding $r_{e}$ independent of magnetic field, $B$.  In the
current work we generalize to keep the $B$-dependence of both $a$ and
$r_e$. This gives $a=a_{bg}(1-\Delta B/(B-B_0))$ and
$r_e=r_{e0}/(1-(B-B_0)/\Delta B)^2$, where
$r_{e0}=-2\hbar^2/ma_{bg}\Delta \mu\Delta B<0$. We thus have a
field-dependent $g_2(B)$ for given values of $B_0$, $\Delta B$,
$\Delta \mu$, and $a_{bg}$. Notice that $r_e=r_{e0}(1-a_{bg}/a)^2$ and
\begin{equation}
  \label{eq:g2(a)}
  g_2(a)=\frac{a^2}{3}-\frac{a r_{e0}}{2}(1-\frac{a_{bg}}{a})^2.
\end{equation}
Hence $g_2$ diverges when $a\to 0$ (referred to as zero-crossing) or
$a\to\infty$ (on resonance).  These are well-known features for model
potentials like the square-well and van der Waals interactions.

\begin{figure}[tbh]
  \epsfig{file=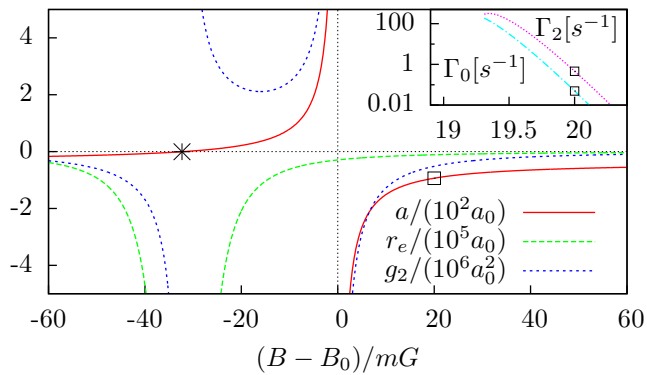,clip=true,scale=1.00}
  \caption{(color online) Scattering length $a$, effective range
    $r_e$, and coupling constant $g_2$ as function of $B$-field for
    the narrow $^{39}$K Feshbach resonance at $B_0=825$~G. The inset
    shows the MQT rate $\Gamma_2$ (and $\Gamma_0$ for $g_2=0$) for
    $N=242$ and $b=1.84\mu$m. The cross and open square are as in
    Fig.~\ref{fig2}.}
  \label{fig3}
\end{figure}

As a concrete example we use the extremely narrow Feshbach resonance
in $^{39}$K found at $B_0=825$G, with $\Delta B=-0.032$G,
$\Delta\mu=3.92\mu_B$, and $a_{bg}=-36a_0$ \cite{errico07}. For this
resonance we find a large $r_{e0}=-2.93\times10^4a_0$.  The variation
of $a$, $r_e$ and $g_2$ as function of $B$ is shown in
Fig.~\ref{fig3}.  We use trap length $b=1.84\mu\textrm{m} =3.48\times
10^4a_0$ in all calculations unless indicated otherwise.

\paragraph{Critical Particle Number Near Zero-crossings}
As noted above, large values of $g_2$ are possible at zero-crossings
($a=0$) since $r_e$ diverges here, see e.g. Fig.~\ref{fig3} around
$B-B_0\sim-32$mG. The scattering length is currently being tuned with
extreme accuracy near such zero-crossings in $^{39}$K around 350G with
control down to the 0.06$a_0$ level \cite{fattori2008}. However, this
is for a broad resonance and the same level of tuning around narrow
ones is likely much more challenging, particularly in high magnetic
fields. Another narrow resonance at $B_0=25.85$G with $\Delta
B=0.47$G exists in $^{39}$K \cite{errico07}. It will show the same
features as the one considered here and the lower magnetic field could
make tuning easier.

The advantage is that many particles $N_c\propto 1/|a|$ can be
accommodated in the condensate.  However, for $a\rightarrow 0$ we have
$ag_2\rightarrow -r_{e0}(a_{bg})^2/2$, i.e. a finite limit.
Remembering that $r_{e0}$ and $\nabla^2|\Psi|^2$ are negative, the
last term in the energy functional eq.~\eqref{efunc} also becomes
negative. Thus larger densities or density fluctuations gives lower
total energy. This implies less stability and smaller $N_c$ near
$a=0$.

\begin{figure}[tbh]
  \epsfig{file=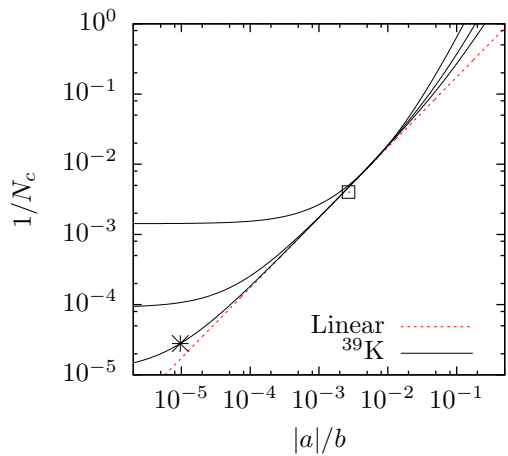,scale=1.00}
  \caption{(color online) Critical particle number $N_c$ as function
    of scattering length $a$. The linear dashed (red) line is the
    well-known $g_2=0$ result $N|a|/b=k_0$. The solid (black) lines
    are for a narrow $^{39}$K resonance and trap length $b=1.84\mu$m,
    (bottom), $0.5b$ (middle), and $0.2b$ (top).  The cross and open
    square are as in Fig.~\ref{fig2}.}
    \label{fig4}
\end{figure}

To calculate quantitative effects on $N_c$ we focus on the critical
line between the unstable and metastable regions in Fig.~\ref{fig2}
for $g_2\le 0$.  We fit the dependency as
\begin{equation}
  \label{eq:Nc-fit}
  N_c=k_0 \frac{b}{|a|} \times \left(1-k_1 \frac{g_2(a)}{b^2}\right)^{-1},
\end{equation}
where $k_1=14.5$, as shown in Fig.~\ref{fig2}. With $g_2(a)$ given by
eq.~\eqref{eq:g2(a)}, $N_c$ becomes a function of $|a|/b$ for fixed
values of $a_{bg}/b$ and $r_{e0}/b$.  In Fig.~\ref{fig4} we plot
$1/N_c$ as function of $|a|/b$ for the $^{39}$K resonance with various
trap lengths.  The curve deviates from the linear $g_2=0$ result both
at $|a|\sim 0$ and $|a|\sim b$.  Squeezing the trap makes the
effect even more pronounced.  In the limit $|a|/b\to 0$ the critical
particle number approaches a finite value, $N_c \to 2 k_0 b^3 /(k_1
|r_{e0}| a_{bg}^2)$.  Large effects also occur near $|a|\sim b$,
however, here the allowed particle number is very small.

To exemplify with the $^{39}$K resonance we use $B-B_0=-32.3$mG (cross
in Fig.~\ref{fig3}) giving $a=-0.334a_0$ and $g_2/b^2=-0.0461$.  The
critical particle number from eq.~\eqref{eq:Nc-fit} is $N_c=3.58\times
10^4$. The values are indicated by crosses in Fig.~\ref{fig2} and
Fig.~\ref{fig4}. The $g_2=0$ prediction is $N_c=5.975\times 10^4$. In
conclusion, we predict 40\% reduction in $N_c$ for macroscopic
particle number $N_c\sim 10^4$.

\paragraph*{Macroscopic Quantum Tunneling}
Close to $a=\infty$, $g_2/b^2$ is still a small term in typical traps.
As the MQT rate \cite{mqt} is exponentially dependent on the integral
under the barrier, we expect changes due to non-zero $g_2$ to be
amplified. A similar point was raised in \cite{ueda1998} where the MQT
rate was shown to increase dramatically close to $N_c$. The MQT rate,
$\Gamma_2$, can be obtained from field theory \cite{stoof1997} by
considering the bounce solution of the effective action in variational
$q$-space with potential $E(q)/N\hbar\omega$.  All MQT calculations
are done within the variational approach, eq.~\eqref{energy}. Since we
assume spherical symmetry, we use $\omega=\bar{\omega}=(\omega_x
\omega_y\omega_z)^{1/3}$ when relating to experiment. For comparison,
we denote by $\Gamma_0$ the rate with $g_2=0$.

We are interested in metastable states with small barriers and large
MQT rates, thus we have to work close to $N_c$ \cite{ueda1998}.  Below
we find that $g_2/b^2$ is only important for the rate when $N$ is of
order $10^2$ or less.  Since the three-body loss depends on the
density to the third power \cite{adhikari2002}, we expect it to be
small in the outer metastable minimum and to be large in the inner
one. Thus, the physical picture is that of a metastable BEC in the
outer minimum that knocks on the barrier as a coherent state with a
common quantum tunneling probability. When it tunnels to the inner
minimum it swiftly decays as the large increase in density amplifies
the three-body loss.  Thermal fluctuations can of course also induce
MQT and we therefore need to operate at very low temperature.  An
estimate of the thermal tunneling rate is given in \cite{stoof1997},
and leads to the following criterion for thermal fluctuations to be
suppressed: $E(q_m)-E(q_0)\gg k_B T$, where $q_m$ and $q_0$ are the
positions of the barrier maximum and the outer minimum, respectively.

For the $^{39}$K resonance we now consider $B-B_0= 20$mG (see
Fig.~\ref{fig3}) where $g_2/b^2=-4.3\times 10^{-4}$ and $a=-93.6a_0$.
The critical particle number becomes $N_c\simeq249$ (see
Fig.~\ref{fig4}).  We now pick $N=242<N_c$ to obtain metastability
(see Fig.~\ref{fig2}).  The rates are found to be $\Gamma_2=0.49$
s$^{-1}$ and $\Gamma_0=0.05$ s$^{-1}$ (shown by the open squares in
the inset of Fig.~\ref{fig3}), i.e. a tenfold enhancement of the MQT
rate. The temperature must be below 8~nK for thermal fluctuations to
be small.  It is possible to obtain larger $N$ with the cost of a
smaller effect on the MQT rate.  However, going to $N\sim 10^{3}$ does
not seem possible.

We propose to start from the $a>0$ side where the condensate is
stable. Preparing a sample with low enough $N$ and temperature of a
few nK for the effect to be observable is the very difficult challenge
for the MQT scenario. A sudden ramp of the magnetic field to the
appropriate $a<0$ is then performed. The system has to retain a small
barrier so that the MQT rate is considerable and to avoid collapse.
Then one monitors $N$ as a function of time, as was done in e.g.
\cite{donley2001}. The low $N$ is also challenging for optical imaging
of the atomic cloud, but should be possible with current techniques.
We expect not only to see modified MQT rates, but also changes in the
decay after the condensate tunnels to a high density state; The
three-body recombination is very sensitive to the density, and the
inner barrier caused by the $g_2$ term should affect this process as
well.

\paragraph*{Other Signatures}
From the structure of the $g_2$ term it is clear that density
variations are needed to see effects. As we have demonstrated already,
tighter traps amplify the contribution, and we expect optical
lattices to do the same.  Solitons and vortices are other features
with density variation. For a simple one vortex state, the
$|\Psi|^2\nabla^2|\Psi|^2$ term will leave the core and the asymptotic
regions unchanged but change the profile in between. Work is in
progress to extract these changes \cite{thoger2009}.

\paragraph*{Conclusion}
We have explored the effect of higher-order terms in the
Gross-Pitaevskii description of a Bose-Einstein condensate,
particularly the effective range correction near a Feshbach resonance.
Using both a variational and numerical approach we find an interesting
new phase diagram for the stability of a condensate with negative
scattering length.

The critical particle number is strongly affected near zero-crossings
of the scattering length for narrow resonances. Effects of 40\% are
found for particle number of order $10^4$.  These deviations increase
for more narrow resonances or tighter traps. We find that the critical
number will be reduced as the zero-crossing is approached from either
side since the higher-order term is always attractive.  
The dipole-dipole interaction is usually important near zero-crossings
but we estimate it to be smaller than the higher-order term for narrow
resonances and tight traps.
Macroscopic
quantum tunneling is also modified by these higher-order corrections
and we have discussed some experimental conditions for exploring the
physics.  Narrow Feshbach resonances are the best way to isolate the
effect of tunneling. However, one needs small samples of order $10^2$
particles and temperatures around 10 nK. Other possible experimental
signatures are tight traps, optical lattices, solitons, and vortices
in rotating BECs.

In a more general sense, the fact that higher-order interactions can
become large and dominant in determining the stability properties
means that it is no longer a correction and that still higher terms
might become important. This is the subject of future work. For the
moment we have demonstrated that experiments targeting the regions
discussed above could find interesting new stability properties beyond
those that were already understood about a decade ago.

\paragraph*{Acknowledgments}
Discussions and encouragement from Aksel S. Jensen is gratefully
appreciated. NTZ acknowledges useful comments from Doron Gazit, as
well as support from the Villum Kann Rasmussen foundation.


\begin{thebibliography}{99}
\bibitem{dalfovo1999} F. Dalfovo, S. Giorgini, L.P. Pitaevskii, and S. Stringari, Rev. Mod. Phys. {\bf 71}, 463 (1999).

\bibitem{ruprecht1995} P.A. Ruprecht, M.J. Holland, K. Burnett, and M. Edwards, Phys. Rev. A {\bf 51}, 4704 (1995).

\bibitem{gammal2001} A. Gammal, T. Frederico, and L. Tomio, Phys. Rev. A {\bf 64}, 055602 (2001).

\bibitem{baym1996} G. Baym and C.J. Pethick, Phys. Rev. Lett. {\bf 76}, 6 (1996).

\bibitem{bradley1997} C.C. Bradley, C.A. Sackett, and R.G. Hulet, Phys. Rev. Lett. {\bf 78}, 985 (1997).

\bibitem{roberts2001} J. L. Roberts {\it et al.}, Phys. Rev. Lett. {\bf 86}, 4211 (2001).

\bibitem{donley2001} E.A. Donley, Nature {\bf 412}, 295 (2001).

\bibitem{lahaye2008} T. Lahaye {\it et al.}, Phys. Rev. Lett. {\bf 101}, 080401 (2008).

\bibitem{errico07} C. D'Errico {\it et al.}, New J. Phys. {\bf 9}, 223 (2007).

\bibitem{roati2007} G. Roati {\it et al.}, Phys. Rev. Lett. {\bf 99}, 010403 (2007).

\bibitem{pet07} A. Collin, P. Massignan, and C.J. Pethick, Phys. Rev. A {\bf 75}, 013615 (2007).

\bibitem{ueda1998} M. Ueda and A.J. Leggett, Phys. Rev. Lett. {\bf 80}, 1576 (1998).

\bibitem{bruun05} G.M. Bruun, A.D. Jackson, and E.E. Kolomeitsev, Phys. Rev. A {\bf 71}, 052713 (2005).


\bibitem{fattori2008} M. Fattori {\it et al.}, Phys. Rev. Lett. {\bf 101}, 190405 (2008).

\bibitem{mqt} A nice introduction to MQT with references is found in D. Mozyrsky {\it et al.}, Phys. Rev. A {\bf 76}, 051601(R) (2007).

\bibitem{adhikari2002} S.K. Adhikari, Phys. Rev. A {\bf 66}, 013611 (2002).

\bibitem{stoof1997} H.T.C. Stoof, J. Stat. Phys. {\bf 87}, 1353 (1997).

\bibitem{thoger2009} M. Th{\o}gersen and N.T. Zinner, in preparation.

\end{thebibliography}
\end{document}